# Theoretical demonstration of the possibility of using binary alloys or solid solutions in ternary systems as geothermometers.

Ya.I. Korepanov[1]

[1] D.S. Korzhinsky Institute of Experimental Mineralogy of the Russian Academy of Sciences (IEM RAS)

*e-mail: yakoff@iem.ac.ru*

Abstract

The investigation of the processes of mineral deposit formation and their history is a fundamental task. Solving this task can increase mining efficiency and make a significant contribution to understanding the formation of the Earth's crust. The main approach to solving this task is determining the values of thermodynamic functions for phases present in deposits. One of the most promising and flexible methods for studying thermodynamic properties is the EMF method. This article describes the basic principles of the EMF method and provides all possible interpretations of the obtained results. It has also been demonstrated that direct EMF measurements can determine the composition of a solid solution or alloy in a multicomponent system if its EMF dependencies are known. Based on this fact, it can be concluded that any alloy or solid solution in a multi-phase association, which undergoes composition changes with temperature and can be quenched, can be used as a geothermometer or geobarometer.

**1. Introduction.**

Investigation of thermodynamic properties of alloys and solid solutions is a fundamental task in both material science and metallurgy [1-6]. It serves as a tool for searching for materials with specified properties. Additionally, it is important in geology [1] for understanding the conditions of mineral deposits formation.

The isobaric potential (Gibbs free energy) is the most convenient thermodynamic function for describing equilibria in multicomponent systems. This function is characteristic for independent variables $P$ and $T$, as it explicitly expresses various properties of the system in terms of its derivatives [5,6]. The EMF method [1,12-25] is one of the most significant and direct methods for studying thermodynamic properties. This article attempts to theoretically substantiate the idea that in EMF cells, the difference in chemical potentials between the comparison system and the sample under study is determined first and foremost. The reaction energy is a consequence of this determined value. The EMF method allows for the accurate determination of the composition of a solid solution in binary or ternary equilibrium. Any alloy or solid solution in a ternary phase association, which changes its composition depending on temperature and is quenched, can be used as a geothermometer.

## 2. Theoretical background.

For the correct and most logical construction of arguments, it is necessary to discuss such things as chemical potential, the principle of operation of an EMF cell, and possible interpretations of the results obtained as the basis for proving the postulated statements.

### 2.1. Chemical potential.

In the thermodynamics, the chemical potential of a component refers to the amount of energy that can be absorbed or released when the concentration of component undergoes a change. In the case of a multi-phase system, the chemical potential of a component is defined as the rate at which the system's free energy changes in response to the addition of atoms belonging to that component. This can be viewed as the partial derivative of the free energy with respect to the quantity of the component, while keeping the concentrations of all other components in the multi-phase system constant. In situations where the pressure and temperature remain constant, and the quantities of components are measured in moles, the chemical potential corresponds to the partial molar Gibbs free energy. At a state of chemical or phase equilibrium [7-8], where the sum of the product of chemical potentials and stoichiometric coefficients [9-11] equals zero, the free energy is minimized.

**A logical condition for equilibrium [10] is the following statement: "In any equilibrium heterogeneous system, the chemical potential of each component is the same in all phases."**

### 2.2. EMF method.

Under the condition of the existence of an ionic conductor with a constant valence and low electronic conductivity, it is possible to create a EMF cell:

(-)A|solid state electrolyte A⁻|AB(solid solution)(or possible ternary phase association A-B-C)(+)

Here A is the reference system, and AB is the sample system. The obtained EMF values, measured either by a compensation circuit or by a voltmeter with a resistance of the order of $10^{12}$ ohms, correspond with an accuracy of up to a coefficient for the work of transferring one mole of substance A through a solid electrolyte:

$$\mu_A^{reference\ system} - \mu_A^{sample\ system} = -z_A FE$$

Where $z_A$- valence of ions in electrolyte, F- Faradays constant, $E$- measured values of EMF. A fully detailed description of the EMF method can be fined in this papers: [1-3,6,12-25].

Consider chemical potential as $\mu_i = \frac{\partial G_i}{\partial N_i} = \mu_i^0 + RTln(a_i)$, where $a_i$-effective concentration i-th component of studying sample in general the main equation will be:

$$RTln\left(a_i^{sample\ system}/a_i^{reference\ system}\right) = -z_A FE$$

Or in the case of our system

$$RTln(a_A^{AB}/a_A^A) = -z_A FE$$

$a_A^A = 1$ by the description, that is, for any properly open cell in which the measurement is made relative to a pure element, the effective concentration of this component is determined.

The main criterion for the reliability of the obtained results is the reproducibility of the measurements and the equilibrium state of the sample under study.

## 2.3. Interpretation of EMF results.

Let's consider a three-component system consisting of components A, B, and C. In this system, there is a triple phase association involving the phases $A_3BC_2$, $A_xB_{1-x}$, and $A_2C$ in such type of the EMF cell:

$$A \mid \text{electrolyte A} \mid A_3BC_2, A_xB_{1-x}, A_2C$$

The relatively pure component A is used as a reference system. According to the definition of the chemical potential $\mu_i = \mu_i^0 + RT\ln(a_i)$, it implies that for any cell of this type, under constant $P$ and $T$ conditions, the "effective concentration" of component A in the sample system is measured. This statement holds true regardless of whether all phases of the sample system are linear.

Assume that a reaction takes place

$$A_3BC_2 + \frac{1}{1-x}A = \frac{1}{1-x}A_xB_{1-x} + 2A_2C \qquad \text{R1}$$

Then the equilibrium state can be written as follows:

$$\Delta G_{A_3BC_2} + \frac{1}{1-x}\Delta G_A = 2\Delta G_{A_2C} + \frac{1}{1-x}\Delta G_{A_xB_{1-x}}$$

The measured value in the cell (by definition chemical. potential)

$$\mu_A^{A_xB_{1-x},A_2C,A_3BC_2} - \mu_A^A = -z_A FE \qquad (1)$$

By the definition

$$\Delta_r G = \sum_{products} \Delta G^0 - \sum_{reagents} \Delta G^0,$$

$$\mu_A^{A_xB_{1-x},A_2C,A_3BC_2} = \mu_A^0 + RT\ln(a),$$

where a is the effective concentration of component A,

$$\Delta G_{phase}^0 = \sum_{i=0}^{n} x_i^{phase} \mu_i$$

$$\Delta G_{A_3BC_2}^0 = 3\mu_A^{A_3BC_2} + \mu_B^{A_3BC_2} + 2\mu_C^{A_3BC_2}$$

and the measured value can be interpreted as follows:

$$\Delta_r G = 2\mu_A^{AC} + \mu_C^{AC} + \frac{1}{1-x}(x\mu_A^{A_xB_{1-x}} + (1-x)\mu_B^{A_xB_{1-x}}) - 3\mu_A^{A_3BC_2} - \mu_B^{A_3BC_2} - 2\mu_C^{A_3BC_2} - \frac{1}{1-x}\mu_A^A \qquad (2)$$

By the condition of equilibrium in a chemical system, the potential of each component for each phase in stable equilibrium is the same. This means that in any equilibrium heterogeneous system, the chemical potential of each component is equal in all phases.

According to the condition of equilibrium, the chemical potential of component A in phase $A_3BC_2$ will be the same as the chemical potential of component A in phases $A_xB_{1-x}$ and $A_2C$.

Similarly, the chemical potential of component B and component C will also be the same in all phases.

$$\mu_i = \mu_i^{A_X B_{1-X}} = \mu_i^{A_2 C} = \mu_i^{A_3 B C_2} \qquad (3)$$

This condition of equal chemical potentials holds true for each component in each phase of the system. It is a fundamental principle of equilibrium in heterogeneous chemical systems.

Thus from equation (1), (2) and (3) obtain:

$$\Delta_r G = \frac{1}{1-x}(\mu_A^{A_X B_{1-X}, A_2 C, A_3 B C_2} - \mu_A^A) = -n\frac{1}{1-x}FE$$

In general, this equation can be written as follows

$$\Delta_r G = m(\mu_A^{ABC, AC, AB} - \mu_A^A) = -z_A m F E$$

ABC+A =AC+AB                R2

For the reaction R2 m=1. But, for example, for a reaction of the form BC$_2$+3A=A$_3$BC$_2$ m=3, and for R1 $m = \frac{1}{1-x}$

Thus, we have obtained that the EMF values can be interpreted as $\Delta_r G$. It also follows from the chemical potential that, under constant $P,T$ conditions, the values obtained can be interpreted as values for an alloy $A_X B_{1-X}$ in equilibrium.

## 3. Theoretical demonstration of the possibility of using binary alloys or solid solutions in ternary systems as a geothermometer.

In any electrochemical cell with an electrolyte conducting component A of a ternary-phase association, the difference in chemical potential of component A is determined first and foremost. This is done by comparing the potentials of component A in the reference system and in the sample system. Therefore, it can be concluded that the effective concentration is directly determined in the electrochemical cell when the measurement is relative to a pure element, or the difference in effective concentrations in other cases. This fact remains true regardless of whether all phases of the sample system are linear or solid solutions. The term "effective concentration" or "activity" accurately represents the fundamental significance of the problem.

Based on the above facts:
1. Any electrochemical cell essentially functions as a "effective concentration" EMF element.
2. The resulting chemical potential difference can be interpreted as the chemical potential of the i-th component in a solid solution A$_x$B$_{1-x}$ at constant pressure and temperature (as there is often a temperature dependence of composition in most systems).
3. The resulting chemical potential difference can be interpreted as a hypothetical reaction that could occur in the sample system in a ternary system under the condition of a closed external circuit.

Point "2" is always satisfied (due to the definition of chemical potential and the equilibrium condition of a closed system), and to prove point "3", it is necessary to conduct a sufficient number of experiments to confirm that such a reaction will indeed occur. In other words, a thorough study of the phase relations in the system under investigation is required.

Therefore, for any equilibrium system that contains a solid solution with known or determinable thermodynamic parameters, and if the solid solution can be quenched, such solid solution can be considered as a geo-(thermo/baro)-meter. This is because its composition, which indicates the effective concentration of the component being studied in the system, directly depends on the temperature and pressure during its formation.

## 4. Discussion.

In some articles, the authors (e.g. [14, 15]) overlook the importance of chemical potential, especially when studying thermodynamic data of linear phases. The fundamental physical meaning of the chemical potential of the component being studied lies in the change in the thermodynamic function with an infinitesimal increase in concentration. This can be expressed as $\mu_i = \frac{\partial G_i}{\partial N_i} = \mu_i^0 + RT ln(a_i)$, where $a_i$ represents the effective concentration of the studied component. In other words, the chemical potential is a value up to a constant that determines the value of the effective concentration of the component being studied. Therefore, in an electrochemical experiment, the focus is not on investigating the reaction itself, contrary to the belief of certain groups of scientists, but rather on determining the ratio of the effective concentration of the component in the reference system and the sample system. This fact sometimes leads to a partial loss of fundamental results, as researchers may mistakenly believe that they are directly investigating the reaction, whereas the reaction is merely an interpretation of the obtained results. Additionally, the presence of phases with variable composition in multiphase associations is sometimes overlooked.

As demonstrated in the article, it is possible to obtain the values of thermodynamic data for a phase with variable composition (solid solutions or alloys) and for linear phases whose thermodynamics research is often in a primary objective from EMF values if there are reliable phase diagram under various P, T conditions. Furthermore, the necessary data for geothermometers and geobarometers (dependence of composition of alloy or solid solution on pressure and temperature) can also be obtained.

*Acknowledgments.*
*This work is fulfilled under Research program № FMUF-2022-0002 of the Korzhinski Institute of Experimental Mineralogy and of*
*leading scientific schools of the Russian Federation within the framework of the project "Chalcogenides: crystal growth, geochemistry, thermodynamics and physical properties" (NSH-2394.2022.1.5).*

Literature.